\documentclass[showpacs,showkeys,amsmath,amssymb,10pt,aps]{revtex4}
\usepackage[T1]{fontenc}
\usepackage[cp1250]{inputenc}
\usepackage{amsfonts}
\usepackage{amsbsy}
\usepackage{amssymb}
\usepackage{amsmath}
\usepackage{bbm}
\usepackage{bm}
\usepackage{epstopdf}
\usepackage{tikz}
\usetikzlibrary{decorations.pathmorphing,decorations.pathreplacing,patterns,calc,arrows,decorations.markings}
\newtheorem{Theorem}{Theorem}
\newtheorem{rem}{Remark}
\newtheorem{pro}{Proposition}

\begin{document}
	
\title[Quantum trajectories for a system interacting with environment in $N$-photon state]{Quantum trajectories for a system interacting with environment in $N$-photon state}

\author{Anita D\k{a}browska$^1$, Gniewomir Sarbicki$^2$, and Dariusz Chru\'sci\'nski$^2$}

\affiliation{$^1$Nicolaus Copernicus University, Toru\'{n}, Collegium Medicum Bydgoszcz, ul. Jagiello\'{n}ska 15, 85-067 Bydgoszcz, Poland}


\affiliation{$^2$Institute of Physics, Faculty of Physics, Astronomy and Informatics,  Nicolaus Copernicus University,
	Grudzi\c{a}dzka 5/7, 87--100 Toru\'n, Poland}


\begin{abstract}
	We derive stochastic master equation for a quantum system interacting with an environment prepared in a continuous-mode $N$-photon state. To determine the conditional evolution of the quantum system depending on continuous in time measurements of the output field the model of repeated interactions and measurements is applied. The environment is defined as an infinite chain of harmonic oscillators which do not interact between themselves and they are prepared initially in an entangled state being a discrete analogue of a continuous-mode $N$-photon state. We provide not only the quantum trajectories but also the analytical formulae for the whole statistics of the output photons and the solution to the master equation. The solution in the  continuous case is represented by a simple diagrammatic technique with very transparent ``Feynman rules''. This technique considerably simplifies the structure of the solution and enables one to find physical interpretation for the solution in terms of a few elementary processes.
\end{abstract}



\maketitle

\section{Introduction}

Together with the increase of experimental techniques of producing propagating wave packets of light of definite numbers of photons and their usage as carriers of quantum information in quantum communication and quantum cryptography \cite{CWSS13,PSZ13,RR15} many theoretical approaches to problem of interaction of such wave packets with a finite-dimensional quantum systems were provided. Pure-state wavefunction approaches \cite{KB16,NKM15} and diagrammatic approaches \cite{SS09,SNA17,RS16} were used to describe the process scattering of $n$ photon packets. Generalized master equations \cite{GEPZ98,DHR02,GJN11,WMSS11,WMHS12,BCBC12,RB17} and stochastic master equations  \cite{GJNC12a,GJN12b,GJN13,GJN14,GZ15,DZA15,SZX16,PDZ16,DSC17,BC17,D18} allowed, for instance, to analyze the problem of excitation of the two-level system interacting with a single as well as with $N>1$ photon packets.

Stochastic master equation (SME) \cite{Car93,BP02,GZ10,WM10}, called also in the literature quantum filtering equation \cite{Bel89,Bel90,BB91}, describes the conditional evolution of an open quantum system interacting with the electromagnetic field depending on the results of continuous in time measurements performed on the output field (the field which carries the information about the system \cite{Bar06,GarCol85}). It is a very useful tool which
can be used to study interaction between light and matter within the rotating wave approximation, quasimonochromatic approximation, and the Markov approximation \cite{GZ10,GCMC18}. The randomness in the model comes from the fact that results of the quantum measurements are in general random. The evolution of the total system consisting of the field and the system is unitary and it is given by quantum stochastic equation of the It\^{o} type \cite{HP84,Par92}. The form of SME depends on the initial state of the field and on the type of the measurement performed on the output field. Solutions to the SMEs are called {\it quantum trajectories} \cite{Car93}. Stochastic master equations for the field in a Gaussian state (vacuum, thermal, coherent, squezeed coherent) are well known and widely used. However, derivation of the SMEs for the case when the field is in a single or in $N>1$ photon state are much more involved than for the Gaussian state. In \cite{GJNC12a,GJN14} the SMEs for the field in a single photon state was derived with making use the cascaded system approach. In \cite{GJN12b,GJN13} to determine the SME for a single-photon state a non-Markovian embedding technique was proposed. The same technique was used in \cite{SZX16} to find the SME for the light in a continuous-mode multi-photon state. The derivation of the SME for the field in Fock input state presented \cite{BC17} was based on a temporal decomposition of the input field (the field before the interaction with the system). All mentioned approaches were formulated in the framework of quantum stochastic calculus of It\^{o} type (QSC) \cite{HP84,Par92} and the input-output formalism \cite{GarCol85}.

In this paper we determine the conditional evolution for an open quantum system interacting with the field prepared in the $N$-photon state. Instead of the methods based on QSC, we use the model of repeated interactions and measurements \cite{AP06,P08,PP09,P10}, known also as a collision model \cite{C17,ACMZ17}. For an application the model of repeated interactions in quantum optics see, for instance, \cite{GCMC18,C17,Doh08,FTVRS17}. We derive the continuous in time evolution of an open quantum system from dynamics generated by a discrete in time sequence of (weak) interactions (collisions) of the system with the bath ``ancillas''. The environment is given as an infinite chain of harmonic oscillators prepared initially in an entangled state being a discrete analogue of a continuous-mode $N$-photon state. We assume that the bath harmonic oscillators do not interact with each other but they interacts one after the other with the quantum system of our interest. After each interaction a measurement is performed on the bath element which has interacted with the system. This approach gives rise to discrete in time stochastic evolution of the system. The initial correlations between the environment elements lead to non-Markovianity of evolution of the quantum system which is not described by a single equation but by a set of coupled equations. We give an intuitive and rigorous interpretation to quantum trajectories. Moreover, we determine the analytical formulae for the whole statistics of the output photons and present the solution to the generalized master equation. {This paper provides generalization of the results for a single-photon state published in \cite{DSC17}.}

Let us stress that in the case when the bath elements are initially uncorrelated and they do not interact between themselves, the model of repeated interactions leads to a Markovian dynamics of an open quantum system and it allows to approximate with arbitrary precision any evolution governed by Gorini-Kossakowski-Linblad-Sudarshan master equation \cite{GKS76,Lin76}. It was shown in \cite{GCMC18,AP06,P08,P10,GG94,B02,GS04,BHJ09} for a Markovian case that the quantum trajectories can be obtained as a continuous limit of discrete quantum trajectories. We prove in the paper that the collision model can be effectively applied also to non-Gaussian state of the input field.

The paper is structured as follows. In Section 2 we present the model of repeated interactions and define an evolution of the compound system and its initial state. Section 3 is devoted to presentation of the conditional state of the system and the future part of the environment which has not interacted with the system. In Section 4 we derive the set of SMEs for the discrete as well as for the continuous case. In Section 5 we present a diagrammatic technique for solving a set of master equations and the corresponding formulae for the statistics of the output photons. Final conclusions are collected in Section 6.

\section{The model of repeated interactions}

We consider a quantum system $\mathcal{S}$ of the Hilbert space $\mathcal{H}_{\mathcal{S}}$ which interacts with an environment consisting of an infinite sequence of harmonic oscillators. We assume that the bath harmonic oscillators do not interact with each other but they interact with the system $\mathcal{S}$ one after the other, each during a time period of the length $\tau$. At a given moment only one of the bath harmonic oscillators is in a contact with $\mathcal{S}$ and each of the harmonic oscillators interacts with $\mathcal{S}$ only once.

The Hilbert space of the bath is given as
\begin{equation}
\mathcal{H}_{\mathcal{E}}=\bigotimes_{k=0}^{+\infty}\mathcal{H}_{\mathcal{E},k},
\end{equation}
where $\mathcal{H}_{\mathcal{E},k}$ stands for the Hilbert space of the $k$-th harmonic oscillator which interacts  with $\mathcal{S}$ in the interval $[k\tau, (k+1)\tau)$. The Hilbert space $\mathcal{H}_{\mathcal{E}}$ can be written as a tensor product
\begin{equation}
\mathcal{H}_{\mathcal{E}}=\mathcal{H}_{\mathcal{E}}^{j-1]}\otimes \mathcal{H}_{\mathcal{E}}^{[j},\;\;\;\mathcal{H}_{\mathcal{E}}^{j-1]}=\bigotimes_{k=0}^{j-1}\mathcal{H}_{\mathcal{E},k},\;\;\;
\mathcal{H}_{\mathcal{E}}^{[j}=
\bigotimes_{k=j}^{+\infty}\mathcal{H}_{\mathcal{E},k}.
\end{equation}
So if $j\tau$ is the current moment then $\mathcal{H}_{\mathcal{E}}^{j-1]}$ is the Hilbert space corresponding to the harmonic oscillators which have already interacted with $\mathcal{S}$  and $\mathcal{H}_{\mathcal{E}}^{[j}$ is the Hilbert space corresponding to the harmonic oscillators which have not interacted with $\mathcal{S}$ yet. Clearly, one has $\mathcal{H}_{\mathcal{E},k}=\mathcal{H}$, where $\mathcal{H}$ is the Hilbert space of the harmonic oscillator.

By $b_{k}$ and $b_{k}^{\dagger}$ we denote operators associated with the $k$-th bath harmonic oscillator defined by
\begin{equation}
b_{k}|N\rangle_{k}=\sqrt{N}|N-1\rangle_{k},
\end{equation}
\begin{equation}
b_{k}^{\dagger}|N\rangle_{k}=\sqrt{(N+1)}|N+1\rangle_{k},
\end{equation}
where $|N\rangle_{k}$ is the number state in $\mathcal{H}_{\mathcal{E},k}$.
They do satisfy the standard canonical commutation relations (CCR)
\begin{equation}
[b_{l}, b_{k}] = 0,\;\;\; [b_{l}^{\dagger}, b_{k}^{\dagger}] = 0,\;\;\;
[b_{l}, b_{k}^{\dagger}] = \delta_{lk}.
\end{equation}
The unitary evolution of the composed $\mathcal{E}+\mathcal{S}$
system describing the repeated interactions up to the time $j\tau$ for $j\geq 1$ is given by \cite{GCMC18,AP06,P08,PP09,P10,C17,FTVRS17,Doh08}
\begin{equation}
U_{j\tau} = \mathbb{V}_{j-1} \mathbb{V}_{j-2} \ldots \mathbb{V}_{0},\;\;\;\;\;U_{0}=\mathbbm{1},
\end{equation}
where the unitary operator $\mathbb{V}_{k}$ acts non-trivially only on $\mathcal{H}_{\mathcal{E},k}\otimes \mathcal{H}_{\mathcal{S}}$, that is,
\begin{equation}\label{intermat}
\mathbb{V}_{k}= \mathbbm{1}_\mathcal{E}^{k-1]} \otimes \mathbbm{V}_{[k}  ,
\end{equation}
and
\begin{equation}\label{}
\mathbbm{V}_{[k} =  \exp\left(-i\tau H_{k}\right)
\end{equation}
with
\begin{equation}\label{hamint}
H_{k} =  \mathbbm{1}_\mathcal{E}^{[k}\otimes H_{\mathcal{S}}+\frac{i}{\sqrt{\tau}}\left(b_{k}^{\dagger}  \otimes \mathbbm{1}_\mathcal{E}^{[k+1}\otimes L-b_{k}\otimes \mathbbm{1}_\mathcal{E}^{[k+1}\otimes L^{\dagger}\right),
\end{equation} 
where $H_\mathcal{S}$ is the Hamiltonian of $\mathcal{S}$ and $L\in\mathcal{B}(\mathcal{H}_S)$. By $\mathcal{B}(\mathcal{H}_S)$ we denote a linear space of bounded operators acting on $\mathcal{H}_{\mathcal{S}}$. A discrete model with the Hamiltonian of the form (\ref{hamint}) one can obtain from the model of interaction of a quantum system with a one-dimensional boson field in the Wigner-Weisskopf approximation with the lower bound of the frequency of the external field extended to $-\infty$. The Hamiltonian $H_{k}$ is derived in the interaction picture eliminating the free evolution of the bath. A detailed discussion about physical assumptions leading to (\ref{hamint}) one can find, for instance, in \cite{GCMC18,C17,FTVRS17}. We set, for simplicity, the Planck constant $\hbar=1$. Clearly, $\mathbb{V}_{0}= \mathbbm{V}_{[0}$. We shall use the Fock representation writing down
\begin{equation}\label{vil}
\exp\left(-i\tau H_{k}\right)  = \sum_{MM^{\prime}} |M\rangle_{k} \langle M^{\prime}|_{k} \otimes \mathbbm{1}_\mathcal{E}^{[k+1} \otimes V_{MM^{\prime}},\ \ \ M,M^{\prime}=0,1,2,\ldots.
\end{equation} 
where $V_{MM^{\prime}} \in\mathcal{B}(\mathcal{H}_S)$. Finally, we assume that the initial state of the composed $\mathcal{E}+\mathcal{S}$ system is the uncorrelated product state vector of the form
\begin{equation}\label{ini0}
|N_{\xi}\rangle\otimes|\psi\rangle,
\end{equation}
where $|\psi\rangle$ is the initial state of $\mathcal{S}$ and $|N_{\xi}\rangle$ is the $N$-photon state of the environment defined as
\begin{equation}\label{nphot}
|N_{\xi}\rangle = \frac{1}{\sqrt{N!}}\left( \mathbf{b}_{\xi}^{\dagger}\right)^{N}|vac\rangle,
\end{equation}
and $|vac\rangle =|0\rangle_{0}\otimes|0\rangle_{1}\otimes|0\rangle_{2}\otimes|0\rangle_{3}\otimes\ldots$
is the vacuum vector in $\mathcal{H}_{\mathcal{E}}$,
\begin{equation}\label{}
\mathbf{b}_\xi^{\dagger} := \sum_{k=0}^{+\infty} \sqrt{\tau}\xi_{k} \mathbf{b}_k^{\dagger} ,
\end{equation}
with
\begin{equation}
\mathbf{b}_{k}^{\dagger}=\mathbbm{1}_\mathcal{E}^{k-1]} \otimes b_{k}^{\dagger}\otimes\mathbbm{1}_\mathcal{E}^{[k+1} .
\end{equation}
Let us note that $\mathbf{b}_{0}^{\dagger}=b_{0}^{\dagger}\otimes\mathbbm{1}_\mathcal{E}^{[1} $. Finally,  $\xi_{k}\in \mathbb{C}$ satisfies $\sum_{k=0}^{+\infty}\tau|\xi_{k}|^2=1$.  
It is clear  that $|N_{\xi}\rangle$ is an entangled state of the bath harmonic oscillators. Some useful properties of $|N_\xi\rangle$ are discussed in \ref{A}. One can check that
\begin{equation}
\mathbf{b}_k |N_{\xi}\rangle=\sqrt{N\tau}\xi_{k}|(N-1)_{\xi}\rangle,
\end{equation}
and hence one gets
\begin{equation}\label{property}
\mathbf{b}_{\xi}|N_{\xi}\rangle=\sqrt{N}|(N-1)_{\xi}\rangle.
\end{equation}
The vector (\ref{nphot}) is a discrete version of the continuous-mode Fock state \cite{BLPS90,L00,O06,RMS07}.

\begin{rem} 
	Let us notice that (\ref{nphot}) does not define the most general $N$-photon state {in $\mathcal{H_{E}}$}. An arbitrary $N$-photon state vector in $\mathcal{H_{E}}$ can be defined by
	\begin{equation}\label{arbitrary}
	|N_{\pmb{\varphi}}\rangle=\frac{1}{\mathcal{N}}\sum_{k_{1},k_{2},\ldots,k_{N}=0}^{+\infty}
	\tau^{N/2}\varphi_{k_{N}\ldots k_{2} k_{1}} \, \mathbf{b}_{k_{N}}^{\dagger}\ldots \mathbf{b}_{k_{2}}^{\dagger} \mathbf{b}_{k_{1}}^{\dagger}|vac\rangle,
	\end{equation}
	where
	${\mathcal{N}}$ stands for the normalization factor. We do not assume any symmetry property for a  tensor $\varphi_{k_{N}\ldots k_{2} k_{1}}$. However, it is clear that the space of $N$-photon states is isomorphic to the space of totally symmetric tensors, that is, tensors $\varphi_{k_{N}\ldots k_{2} k_{1}}$ invariant with respect to an arbitrary permutation of indices $\{k_{N},\ldots, k_{2}, k_{1}\}$. Particular situation corresponds to
	\begin{equation}\label{}
	\varphi_{k_{N}\ldots k_{2} k_{1}} = \xi^{(N)}_{k_N} \ldots \xi^{(1)}_{k_1} ,
	\end{equation}
	where $\xi^{(i)}_{k_i}$ are profiles of $N$ indistinguishable photons. In this paper we consider the simplest scenario when all profiles are the same.
\end{rem}

\section{Repeated measurements and conditional state}

{We assume that after each interaction the measurement is performed on the element of the bath chain just after its interaction with $\mathcal{S}$. The goal of this Section is to present a structure of the conditional state of the compound system depending on the results of the measurement performed on the bath elements. We consider in the paper the measurement of the bath observable
	\begin{equation}\label{obs}
	N_{k}=b_{k}^{\dagger}b_{k}, \;\;\; k=0,1,\ldots.
	\end{equation}
	By $\pmb{\eta}_j$ we denote the stochastic vector $\pmb{\eta}_j = (\eta_j,\eta_{j-1},\ldots,\eta_1)$ representing results of all measurements of (\ref{obs}) obtained up to time $j\tau$.  A sketch of the repeated interactions and measurements model is shown in Fig. 1. Note that in general the process of detection can be delayed.
	\hskip -0.6cm
	\begin{figure}
		\begin{tikzpicture}
		\pgfmathsetmacro\w{10.3};
		\draw[line width=0.8pt,fill=white] (-1.5,.5) rectangle (0,1.5);
		\draw[line width=0.8pt,fill=white] (\w,.5) rectangle ({\w+1.8},1.5);
		\draw[->,>=latex,line width=0.8pt,decorate, decoration={snake,amplitude=.2mm,segment length=2mm}] (0,1) -- (\w,1);
		\draw[->,>=latex] (-1,-1) -- ({\w+1},-1);
		\fill[black,opacity=.5] ({\w/2},1) circle (.2);
		\draw ({\w/2},-.9) -- ({\w/2},-1.1);
		\draw (0,1.5) -- (0,2.1);
		\draw (\w,1.5) -- (\w,2.1);
		\draw[->,>=latex] ({\w/2},2.5) -- ({\w/2},1.6);
		\draw[<->,>=latex] (0,2) -- ({\w/2},2);
		\draw[<->,>=latex] (\w,2) -- ({\w/2},2);
		\node at ({\w/2},.5) {\scriptsize $\mathcal{H}_\mathcal{S} \! \otimes \! \mathcal{H}_{\mathcal{E}\!,k}$};
		\node at ({\w/2},-1.3) {$0$};
		\node at ({\w+1},-1.3) {$x$};
		\node at (-.75,1) {Source};
		\node at (\w+.9,1) {Detector};
		\node at ({\w/2},2.7) {\textit{interaction}};
		\node at ({\w/4},2.2) {\textit{free evolution}};
		\node at ({3*\w/4},2.2) {\textit{free evolution}};
		\pgfmathsetmacro\n{3};
		\foreach \x in {0,1,...,\n} 	{ \draw ({\w/2*\x/(\n+.5},-1.1) -- ({\w/2*\x/(\n+.5)},1.1);  \draw ({\w-\w/2*\x/(\n+.5},-1.1) -- ({\w-\w/2*\x/(\n+.5)},1.1);}
		\node at (\w/2,1.5) { \scriptsize $k$ };
		\foreach \x in {1,2,...,\n}	
		{
			\node at ({\w/2-\w/2*(\x)/(\n+.5},1.5) {\scriptsize $k\!+\!\x$};
			\node at ({\w/2-\w/2*(\x)/(\n+.5},.5) {\scriptsize $\mathcal{H}_{\mathcal{E},k+\x}$};
		}
		\foreach \x in {1,2,...,\n}
		{
			\node at ({\w/2+\w/2*(\x)/(\n+.5},1.5) {\scriptsize $k\!-\!\x$};
			\node at ({\w/2+\w/2*(\x)/(\n+.5},.5) {\scriptsize	$\mathcal{H}_{\mathcal{E},k-\x}$};
		}
		\pgfmathsetmacro\m{2*\n+1};
		\foreach \x in {-\n,...,\n}
		{
			\begin{scope}[shift={({\w/2+\x*\w/(2*\n+1)},-.5)}]
			\draw plot[smooth,domain=-.4:.4] (\x, {4*\x*\x});
			\draw (-.02^.5,4*.02) -- (.02^.5,4*.02);
			\draw (-.06^.5,4*.06) -- (.06^.5,4*.06);
			\draw (-.1^.5,4*.1) -- (.1^.5,4*.1);
			\end{scope}
		}
		\draw [line width = 0.8,decorate,decoration={brace,amplitude=4,mirror}] ({\w/2-1.5*\w/(2*\n+1)},-1.2) -- ({\w/2-.5*\w/(2*\n+1)},-1.2) node [midway,yshift=-10] {\footnotesize $c\tau$};
		\end{tikzpicture}
		\caption[]{A repeated interactions and measurements model with a delayed process of detection}
	\end{figure}
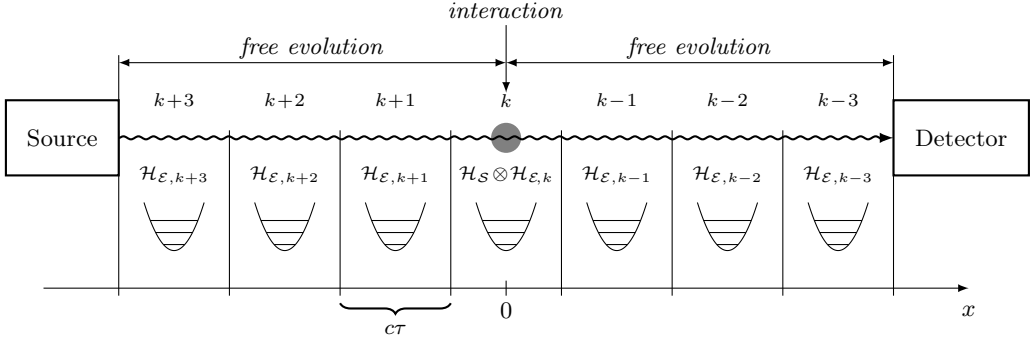

	\begin{Theorem} The conditional state vector of $\mathcal{S}$ and the part of the environment which has not interacted with $\mathcal{S}$ up to $j\tau$ for the initial state (\ref{ini0}) and the measurement of (\ref{obs}) at the moment $j\tau$ is given by
		\begin{equation}\label{cond1}
		|\tilde{\Psi}_{j| \pmb{\eta}_j}\rangle = \frac{|\Psi_{j| \pmb{ \eta}_j}\rangle}{\sqrt{\langle\Psi_{j| \pmb{\eta}_j}|\Psi_{j| \pmb{\eta}_j}\rangle}} ,
		\end{equation}	
		where the unnormalized conditional state vector $|\Psi_{j| \pmb{ \eta}_j}\rangle \in \mathcal{H}_\mathcal{E}^{[j} \otimes \mathcal{H}_\mathcal{S}$ has the following
		structure	
		\begin{equation}\label{cond5}
		|\Psi_{j|\pmb{\eta}_j}\rangle=
		\sum_{M=0}^{N}|M_{\xi}\rangle_{[j,+\infty)}\otimes|\psi_{j|\pmb{\eta}_j}^{M}\rangle .
		\end{equation}
		Moreover,  $|M_{\xi}\rangle_{[j,+\infty)}$ is the unnormalized vector from $\mathcal{H}_{\mathcal{E}}^{[j}$ given by
		\begin{equation}
		|M_{\xi}\rangle_{[j,+\infty)} = \frac{1}{\sqrt{M!}}\left(\sqrt{\tau}\xi_{j}b_{j}^{\dagger}\otimes\mathbbm{1}_\mathcal{E}^{[j+1}+\sum_{k=j+1}^{+\infty}\mathbbm{1}_\mathcal{E}^{[j,k-1]} \otimes \sqrt{\tau}\xi_{k} b_{k}^{\dagger}\otimes\mathbbm{1}_\mathcal{E}^{[k+1}\right)^{M}|vac\rangle_{[j,+\infty)},
		\end{equation}
		where $|vac\rangle_{[j,+\infty)} =|0\rangle_{j}\otimes|0\rangle_{j+1}\otimes\ldots$, and
		the conditional vectors $|\psi_{j|\pmb{\eta}_j}^{0}\rangle$, $|\psi_{j|\pmb{\eta}_j}^{1}\rangle$,\ldots, $|\psi_{j|\pmb{\eta}_j}^{N}\rangle$ from $\mathcal{H}_{\mathcal{S}}$ satisfy the set of coupled recurrence equations
		\begin{equation}\label{rec}
		|\psi_{j+1|\pmb{\eta}_{j+1}}^{M}\rangle = \sum_{M^{\prime}=0}^{N-M} \sqrt{M+M^{\prime} \choose M^{\prime}} \left(\sqrt{\tau} \xi_j\right)^{M^{\prime}} V_{\eta_{j+1} M^{\prime}} |\psi_{j|\pmb{\eta}_j}^{M+M^{\prime}}\rangle,
		\end{equation}
		where the operators $V_{\eta_{j+1} M^{\prime}}\in\mathcal{B}(\mathcal{H}_S)$ are defined in  (\ref{vil}), and initially $|\psi^{N}_{j=0}\rangle=|\psi\rangle$, and $|\psi^{M}_{j=0}\rangle=0$ for $0\leq M\leq N-1$.
	\end{Theorem}
	We use the notation $|0_{\xi}\rangle_{[j,+\infty)}=|vac\rangle_{[j,+\infty)}$. For the proof see \ref{C}.}

The formula (\ref{cond5}) clearly shows that $|{\Psi}_{j| \pmb{\eta}_j}\rangle$ is the entangled state vector of $\mathcal{S}$ and the part of environment which has not interacted with $\mathcal{S}$ yet. The Schmidt rank of  $|{\Psi}_{j| \pmb{\eta}_j}\rangle$ is upper-bounded by the total number of photons `$N$'.
Note, that `$N+1$' vectors $|M_{\xi}\rangle_{[j,+\infty)}$ are mutually orthogonal for different `$M$'. However, the system vectors $|\psi_{j|\pmb{\eta}_j}^{M}\rangle$ are in general not mutually orthogonal. Hence, (\ref{cond5}) does not provide the Schmidt decomposition.
Clearly, initially the state is separable and its Schmidt rank equals 1. The state vector  $|{\Psi}_{j| \pmb{\eta}_j}\rangle$ has a clear physical interpretation: it represents a superposition of  $N+1$ possible scenarios: the future part of the environment can be in the vacuum state $|vac\rangle_{[j,+\infty)}$ or in one of the states $|M_{\xi}\rangle_{[j,+\infty)}$ for $1\leq M\leq N$. Accumulating results of all measurement we gradually eliminate the successive scenarios starting from the possibility of storing $N$ photons in the future and eventually finding the environment in the vacuum state. So sooner or later the state $|\tilde{\Psi}_{j| \pmb{\eta}_j}\rangle$ becomes separable.

Now, performing a partial trace of $|\tilde{\Psi}_{j| \pmb{\eta}_j}\rangle\langle|\tilde{\Psi}_{j| \pmb{\eta}_j}|$ w.r.t. $\mathcal{S}$ we obtain the conditional state of the future part of the environment
\begin{equation}\label{cond3}
\varrho_{j}^{\rm field}=\frac{
	\sum_{M=0}^{N}\sum_{M^{\prime}=0}^{N}\langle\psi^{M^{\prime}}_{j|  \pmb{\eta}_{j}}|\psi_{j|  \pmb{\eta}_{j}}^{M}\rangle|M_{\xi}\rangle_{[j,+\infty)}
	\langle M^{\prime}_{\xi}|_{[j,+\infty)}
}{ \sum_{M=0}^{N} p_j^M\, \langle\psi^{M}_{j|  \pmb{\eta}_{j}}|\psi_{j|  \pmb{\eta}_{j}}^{M}\rangle },
\end{equation}
where
\begin{equation}\label{}
p_j = \sum_{k=j}^{+\infty}\tau|\xi_{k}|^2 .
\end{equation}
Taking into account formula (\ref{norm}) one can check that the probability that at time  $j\tau$ the environment stores in the future $M$ photons for $0\leq M\leq N$ reads
\begin{equation}\label{prob1}
P^{\rm field}_j(M) = \frac{ p_j^M \, \langle\psi^{M}_{j|  \pmb{\eta}_{j}}|\psi_{j|  \pmb{\eta}_{j}}^{M}\rangle} {\sum_{M^{\prime}=0}^{N} p_j^{M^{\prime}}\, \langle\psi^{M^{\prime}}_{j|  \pmb{\eta}_{j}}|\psi_{j|  \pmb{\eta}_{j}}^{M^{\prime}}\rangle} . 
\end{equation}
The above expressions depend on the initial state of $\mathcal{S}$, the photon profile $\xi_{k}$, and the results of all measurement performed up to $j\tau$ indicated by the vector $\pmb{\eta}_{j}$. We can say that the knowledge of the results of the measurement changes our knowledge of the future state of the bath {(this part of the bath which will interact with $\mathcal{S}$ in the future after time $j\tau$)}.

Let us note that the conditional probability of detecting $M$ photons at moment $(j+1)\tau$ when the conditional state of $\mathcal{S}$ and the future part of the environment at $j\tau$ was $|\tilde{\Psi}_{j| \pmb{\eta}_j}\rangle$ is defined as
\begin{equation}
p_{j+1}\left(M \Big|\,|\tilde{\Psi}_{j| \pmb{\eta}_j}\rangle\right)=\frac{\langle{\Psi}_{j| \pmb{\eta}_j}|  \mathbbm{V}_{[j}^{\dagger}\left(|M\rangle_{j}\langle M|_{j}\otimes \mathbbm{1}_\mathcal{E}^{[j+1}\otimes \mathbbm{1}_\mathcal{S}\right) \mathbbm{V}_{[j}  |{\Psi}_{j| \pmb{\eta}_j}\rangle}{\langle{\Psi}_{j| \pmb{\eta}_j}|{\Psi}_{j| \pmb{\eta}_j}\rangle}.
\end{equation}
{Expanding (\ref{vil}) in the Taylor series one finds that $V_{MM^{\prime}}=O(\sqrt{\tau}^{|M-M^{\prime}|})$, where $O(\cdot)$ is the Landau symbol}. Using this result and formula (\ref{cond2}) one  readily checks that
\begin{equation}
p_{j+1}\left(0\Big|\,|\tilde{\Psi}_{j| \pmb{\eta}_j}\rangle\right)=1+O(\tau),
\end{equation}
and for all $M> 0$
\begin{equation}
p_{j+1}\left(M\Big|\,|\tilde{\Psi}_{j| \pmb{\eta}_j}\rangle\right)=O(\tau^M).
\end{equation}
It is, therefore, clear that the probability of detecting more than one photon behaves like  $O(\tau^2)$. In the continuous time limit when $\tau\to dt$ the probability of detecting more that one photon in the time interval of the length $dt$ vanishes. Now  neglecting all terms of order higher than one in $\tau$ and the processes of detecting more that one photon we get from (\ref{rec}) the set of $N+1$ difference equations of the form
\begin{equation}\label{rec2a}
|\psi_{j+1|\pmb{\eta}_{j+1}}^{N}\rangle = V_{\eta_{j+1} 0} |\psi_{j|\pmb{\eta}_j}^{N}\rangle,
\end{equation}
and for  $0 \leq M < N-1$

\begin{eqnarray}\label{rec2b}
|\psi_{j+1|\pmb{\eta}_{j+1}}^{M}\rangle  &=& V_{\eta_{j+1} 0} |\psi_{j|\pmb{\eta}_j}^{M}\rangle+\sqrt{(M+1)\tau} \xi_j V_{\eta_{j+1} 1} |\psi_{j|\pmb{\eta}_j}^{M+1}\rangle,
\end{eqnarray}
where
\begin{eqnarray}\label{vmatrix}
V_{00}&=& \mathbbm{1}_{\mathcal{S}} - i\tau H_{\mathcal{S}} - \tau \frac{1}{2}L^\dagger L + O(\tau^{2}) ,\nonumber\\
V_{10}&=&\sqrt{\tau} L + O(\tau^{3/2}) ,\nonumber\\
V_{01}&=&- \sqrt{\tau} L^\dagger + O(\tau^{3/2}),\nonumber\\
V_{11}&=& \mathbbm{1}_{\mathcal{S}} + O(\tau).
\end{eqnarray}
Note that for the difference equations we have now only $\eta_{j+1}=\{0,1\}$. It is seen from Eqs. (\ref{rec2a}) and (\ref{rec2b}) that due to interaction with the environment prepared in $|N_{\xi}\rangle$ the system $\mathcal{S}$ can emit or absorb at most one photon. The processes of emission or absorption of more than one photon are not considered because their probabilities are $O(\tau^2)$ and they can be ignored.

\section{Stochastic master equation}

The entanglement between $\mathcal{S}$ and the environment makes the evolution of $\mathcal{S}$ non-Markovian. We shall show that a recurrence procedure for the state of $\mathcal{S}$ is given not by a single equation but by a set of coupled equations. {Difference and differential stochastic master equations as well as corresponding master equations are determined in this Section.}

\subsection{Difference stochastic master equations}

To obtain the reduced conditional state of $\mathcal{S}$ one has to take the partial trace of $|\tilde{\Psi}_{j| \pmb{\eta}_j}\rangle \langle\tilde{\Psi}_{j| \pmb{\eta}_j}|$ over the environment. It is easy to check that the conditional state of $\mathcal{S}$ at the time $j\tau$ has the form
\begin{equation}\label{condS}
\tilde{\rho}_{j|  \pmb{\eta}_{j}}
=\frac{\rho_{j|  \pmb{\eta}_{j}}}{\mathrm{Tr}_{\mathcal{S}}\rho_{j|  \pmb{\eta}_{j}}},
\end{equation}
where
\begin{equation}\label{cond4}
\rho_{j|  \pmb{\eta}_{j}}=\sum_{M=0}^{N}  p_j^M\, |\psi^{M}_{j|  \pmb{\eta}_{j}}\rangle\langle\psi_{j|  \pmb{\eta}_{j}}^{M}| .
\end{equation}
The operator $\tilde{\rho}_{j|  \pmb{\eta}_{j}}$ defines the {\it a posteriori} state of $\mathcal{S}$ depending on the results of all measurements performed {on the bath elements after their interaction with $\mathcal{S}$} up to $j\tau$. The quantity
\begin{equation}\label{probtra}
\mathrm{Tr}_{\mathcal{S}}\rho_{j|  \pmb{\eta}_{j}}=\sum_{M=0}^{N} p_j^M\, \langle\psi^{M}_{j|  \pmb{\eta}_{j}}|\psi_{j|  \pmb{\eta}_{j}}^{M}\rangle 
\end{equation}
is the probability of a particular trajectory.

We shall derive a difference stochastic equation for $\tilde{\rho}_{j|  \pmb{\eta}_{j}}$, that is, we shall provide a recipe for the conditional state of $\mathcal{S}$ at the time $(j+1)\tau$ depending on the conditional state of $\mathcal{S}$ at $j\tau$ and the random result of the successive measurement. To simplify our notation we drop the condition $\pmb{\eta}_j$ from now on. Let us introduce the following system operators
\begin{equation}
\rho^{M,M^{\prime}}_{j}=\frac{\mathrm{Tr}_{{\mathcal{E}}^{[j}}\left[\left(b_{j}^{N-M}\otimes \mathbbm{1}_\mathcal{E}^{[j+1}\otimes\mathbbm{1}_\mathcal{S}\right)|\Psi_{j}\rangle\langle\Psi_{j}|\left(\left(b_{j}^{\dagger} \right)^{N-M^{\prime}}\otimes
	\mathbbm{1}_\mathcal{E}^{[j+1}\otimes\mathbbm{1}_\mathcal{S}\right)
	\right]}{\tau^{(2N-M-M^{\prime})/2}\xi_{j}^{N-M}(\xi_{j}^{\ast})^{N-M^{\prime}}},
\end{equation}
where $ 0\leq M\leq N$ and $ 0\leq M^{\prime}\leq N$. Of course, $\rho^{N,N}_{j}=\rho_{j}$ and  $\rho^{M,M^{\prime}}_{j}=\left(\rho_{j}^{M^{\prime},M}\right)^{\dagger}$. 

Let
\begin{equation}\label{oper}
\tilde{\rho}_{j}^{M,M^{\prime}}=\frac{\rho_{j}^{M,M^{\prime}}}{\mathrm{Tr}_{\mathcal{S}}\rho_{j}}.
\end{equation}
It is clear that $\tilde{\rho}_{j}^{N,N}=\tilde{\rho}_{j}$. Making use of the property
\begin{equation}
b_{j}\otimes\mathbbm{1}_\mathcal{E}^{[j+1} |M_{\xi}\rangle_{[j,+\infty)}=\sqrt{\tau M}\xi_{j}|(M-1)_{\xi}\rangle_{[j,+\infty)}
\end{equation}
one can easily check that initially
\begin{equation}\label{ini}
\tilde{\rho}_{j=0}^{M,M^{\prime}}=\frac{N!}{\sqrt{M!M^{\prime}!}}\, \delta_{MM^{\prime}}\,|\psi\rangle\langle\psi|.
\end{equation}

\begin{pro} If  $\eta_{j+1}=0$, then 
	
	\begin{eqnarray}\label{filter1d}
	\tilde{\rho}_{j+1}^{M,M^{\prime}} &=& \tilde{\rho}_{j}^{M,M^{\prime}} + \tau \Big( \tilde{\rho}_{j}^{M,M^{\prime}}k_{j}  -i[H_{\mathcal{S}},\tilde{\rho}_{j}^{M,M^{\prime}}] - \frac{1}{2}\left\{L^{\dagger}L,\tilde{\rho}_{j}^{M,M^{\prime}}\right\} \nonumber\\
	&-&\tilde{\rho}_{j}^{M,M^{\prime}-1}L\xi_{j}^{\ast} - L^{\dagger}\tilde{\rho}_{j}^{M-1,M^{\prime}}\xi_{j} - \tilde{\rho}_{j}^{M-1,M^{\prime}-1}|\xi_{j}|^2\Big)  + O(\tau^2) ,
	\end{eqnarray}
	where
	
	\begin{equation}\label{intensity1}
	k_{j} = \mathrm{Tr}_{\mathcal{S}}\left( L^{\dagger}L\tilde{\rho}_{j}
	+ \xi_{j}^{\ast}L\tilde{\rho}_{j}^{N,N-1}+ \xi_{j}
	\tilde{\rho}_{j}^{N-1,N}L^{\dagger}+|\xi_{j}|^2\tilde{\rho}_{j}^{N-1,N-1} \right) .
	\end{equation}
	If  $\eta_{j+1}=1$, then
	\begin{equation}\label{filter1h}
	\tilde{\rho}_{j+1}^{M,M^{\prime}} = \frac{1}{k_{j}}\left(L\tilde{\rho}_{j}^{M,M^{\prime}}L^{\dagger}+\xi_{j}^{\ast}L\tilde{\rho}_{j}^{M,M^{\prime}-1}+\xi_{j}\tilde{\rho}_{j}^{M-1,M^{\prime}}L^{\dagger}+|\xi_{j}|^2\tilde{\rho}_{j}^{M-1,M^{\prime}-1}\right)+ O(\tau).
	\end{equation} 
\end{pro}

Proof: The derivation of the difference equation for $\tilde{\rho}_{j}$ we start from the determination of the difference equations for the unnormalized operators $\rho_{j}^{M,M^{\prime}}$. One can check, referring to Eqs. (\ref{rec2a}) and (\ref{rec2b}), and the relation
\begin{equation}
\| |M_{\xi}\rangle_{[j+1,+\infty)}\|^2 =\||M_{\xi}\rangle_{[j,+\infty)}\|^2 -M\tau|\xi_{j}|^2 \||(M-1)_{\xi}\rangle_{[j,+\infty)}\|^2 + O(\tau^2) ,
\end{equation}
where $\|\cdot\|^2=\langle \cdot|\cdot \rangle$, that when the result of the measurement at $(j+1)\tau$ is zero, that is, $\eta_{j+1}=0$,  then at $(j+1)\tau$ we have
\begin{eqnarray}\label{filter1b}
\rho_{j+1}^{M,M^{\prime}} &=& \rho_{j}^{M,M^{\prime}}-\tau \Big(i[H_{\mathcal{S}},\rho_{j}^{M,M^{\prime}}]+\frac{1}{2}\left\{L^{\dagger}L,\rho_{j}^{M,M^{\prime}}\right\}\nonumber\\
&+&\rho_{j}^{M,M^{\prime}-1}L\xi_{j}^{\ast}+L^{\dagger}\rho_{j}^{M-1,M^{\prime}}\xi_{j}+ \rho_{j}^{M-1,M^{\prime}-1}|\xi_{j}|^2 \Big) + O(\tau^2),
\end{eqnarray}
where $0\leq M\leq N$ and $0\leq M^{\prime}\leq N$. In particular, for $\rho_{j+1}$ we get
\begin{eqnarray}\label{filter1a}
\rho_{j+1} &=& \rho_{j}-\tau \Big(i[H_{\mathcal{S}},\rho_{j}]+\frac{1}{2}\left\{L^{\dagger}L,\rho_{j}\right\}\nonumber\\
&+&\rho_{j}^{N,N-1}L\xi_{j}^{\ast}+L^{\dagger}\rho_{j}^{N-1,N}\xi_{j}+ \rho_{j}^{N-1,N-1}|\xi_{j}|^2 \Big) + O(\tau^2).
\end{eqnarray}
The conditional probability of detecting zero photons at the moment $(j+1)\tau$ when the conditional state of $\mathcal{S}$ at $j\tau$ was $\tilde{\rho}_{j}$ is defined by
\begin{equation}
p_{j+1}(0|\tilde{\rho}_{j})=\frac{\mathrm{Tr}_{\mathcal{S}}\rho_{j+1}}{\mathrm{Tr}_{\mathcal{S}}\rho_{j}}
\end{equation}
with $\rho_{j+1}$ given by Eq. (\ref{filter1a}).
One can readily find that
\begin{equation}
p_{j+1}(0|\tilde{\rho}_{j})=1- k_{j}\tau+O(\tau^2) .
\end{equation}
In the next step,
using the property
\begin{equation}
\frac{1}{\mathrm{Tr}_{\mathcal{S}}\rho_{j+1}} =\frac{1}{ \mathrm{Tr}_{\mathcal{S}}\rho_j  }    \left(1+k_{j} \tau \right) + O(\tau^2) ,
\end{equation}
we get from (\ref{filter1b}) the set of difference equations (\ref{filter1d}). 

Now let us consider the case when the result of the measurement at $(j+1)\tau$ is one, that is, $\eta_{j+1}=1$.  Then from Eqs. (\ref{rec2a}) and (\ref{rec2b}) we obtain for $\rho_{j}^{M,M^{\prime}}$ the recurrence formula
\begin{equation}\label{filter1f}
\rho_{j+1}^{M,M^{\prime}} = \tau\left(L\rho_{j}^{M,M^{\prime}}L^{\dagger}+\xi_{j}^{\ast}L\rho_{j}^{M,M^{\prime}-1}+\xi_{j}\rho_{j}^{M-1,M^{\prime}}L^{\dagger}+|\xi_{j}|^2\rho_{j}^{M-1,M^{\prime}-1}\right)+ O(\tau^2).
\end{equation}
Thus for the unnormalized conditional density matrix $\rho_{j}$ we get
\begin{equation}\label{filter1e}
\rho_{j+1} = \tau\left(L\rho_{j}L^{\dagger}+\xi_{j}^{\ast}L\rho_{j}^{N,N-1}+\xi_{j}\rho_{j}^{N-1,N}L^{\dagger}+|\xi_{j}|^2\rho_{j}^{N-1,N-1}\right)+ O(\tau^2).
\end{equation}
The conditional probability of the outcome one at the moment $(j+1)\tau$ provided that
the conditional state of $\mathcal{S}$ at $j\tau$ was $\tilde{\rho}_{j}$
is given as
\begin{equation}
p_{j+1}(1|\tilde{\rho}_{j})=\frac{\mathrm{Tr}_{\mathcal{S}}\rho_{j+1}}{\mathrm{Tr}_{\mathcal{S}}\rho_{j}},
\end{equation}
where $\rho_{j+1}$ is specified by Eq. (\ref{filter1e}). One can easily check that
\begin{equation}
p_{j+1}(1|\tilde{\rho}_{j})=k_{j}\tau + O(\tau^2) .
\end{equation}
Hence, when the result is one at $(j+1)\tau$, we get (\ref{filter1h}). \hfill $\Box$

\begin{rem} In particular, for the {\it a posteriori} state of $\mathcal{S}$ at $(j+1)\tau$ we have: if $\eta_{j+1}=0$
	\begin{eqnarray}\label{filter1c}
	\tilde{\rho}_{j+1} &=& \tilde{\rho}_{j}+ \tau \Big( \tilde{\rho}_{j}k_{j} 
	-i[H_{\mathcal{S}},\tilde{\rho}_{j}] - \frac{1}{2}\left\{L^{\dagger}L,\tilde{\rho}_{j}\right\} \nonumber\\
	&-&\tilde{\rho}_{j}^{N,N-1}L\xi_{j}^{\ast} -L^{\dagger}\tilde{\rho}_{j}^{N-1,N}\xi_{j} - \tilde{\rho}_{j}^{N-1,N-1}|\xi_{j}|^2\Big)  + O(\tau^2) ,
	\end{eqnarray}
	and if $\eta_{j+1}=0$
	
	\begin{equation}\label{filter1g}
	\tilde{\rho}_{j+1} = \frac{1}{k_{j}}\left(L\tilde{\rho}_{j}L^{\dagger}+\xi_{j}^{\ast}L\tilde{\rho}_{j}^{N,N-1}+\xi_{j}
	\tilde{\rho}_{j}^{N-1,N}L^{\dagger}+|\xi_{j}|^2\tilde{\rho}_{j}^{N-1,N-1}\right)+ O(\tau) .
	\end{equation}
\end{rem}

Let us introduce now the stochastic discrete process
\begin{equation}
n_{j}=\sum_{k=1}^{j}\eta_{k},
\end{equation}
with the increment given by $\Delta n_{j}=n_{j+1}-n_{j}=\eta_{j+1}$. Note that the conditional mean value of $\Delta n_{j}$ is
\begin{equation}\label{mean}
\mathbbm{E}[\Delta n_{j}|\tilde{\rho}_{j}]=k_{j}\tau+O(\tau^2).
\end{equation}
Now, by combining Eqs. (\ref{filter1c}) and (\ref{filter1g}), we find for the {\it a posteriori} state of $\mathcal{S}$ the difference stochastic equation
\begin{eqnarray}\label{filteri}
\tilde{\rho}_{j+1}&=& \tilde{\rho}_{j}+\tau\Big(-i[H_{\mathcal{S}},\tilde{\rho}_{j}]-\frac{1}{2}\left\{L^{\dagger}L,\tilde{\rho}_{j}\right\}
+L\rho_{j}L^{\dagger}\nonumber\\
&+& [\tilde{\rho}_{j}^{N-1,N},L^{\dagger}]\xi_{j}
+[L, \tilde{\rho}_{j}^{N,N-1}]\xi^{\ast}_{j}\Big) \nonumber\\
&+& \bigg\{\frac{1}{k_{j}}\left(
L\tilde{\rho}_{j}L^{\dagger}+L\tilde{\rho}_{j}^{N,N-1}\xi^{\ast}_{j}+\tilde{\rho}_{j}^{N-1,N}L^{\dagger}\xi_{j}\right.\nonumber\\
&+& \left.\tilde{\rho}_{j}^{N-1,N-1}|\xi_{j}|^2\right)
-\tilde{\rho}_{j}\bigg\}\left(\Delta n_{j}-k_{j}\tau\right)
\end{eqnarray}
with the initial conditions: $\tilde{\rho}_{j=0}=|\psi\rangle\langle\psi|$, $\tilde{\rho}_{j=0}^{N,N-1}=\tilde{\rho}_{j=0}^{N-1,N}=0$, and $\tilde{\rho}_{j=0}^{N-1,N-1}=N|\psi\rangle\langle\psi|$. It is clear that when $\Delta n_{j}=0$ (the result of the measurement is zero) then (\ref{filteri}) reduces to (\ref{filter1c}) and when $\Delta n_{j}=1$ (the result of the measurement is one) then all terms proportional to $\tau$ in (\ref{filteri}) are negligible and (\ref{filteri}) becomes equivalent to (\ref{filter1g}). Of course, in order to determine the {\it a posteriori} state of $\mathcal{S}$ at any time $j\tau$ where $j> 0$ we need to use a set of coupled equations depending on the stochastic trajectory up to time $j\tau$. From Eqs. (\ref{filter1d}) and (\ref{filter1h}) we obtain the set of difference stochastic equations  of the form
\begin{eqnarray}\label{filterj}
\tilde{\rho}_{j+1}^{M,M^{\prime}}&=& \tilde{\rho}_{j}^{M,M^{\prime}}+\tau\Big(-i[H_{\mathcal{S}},\tilde{\rho}_{j}^{M,M^{\prime}}]-\frac{1}{2}\left\{L^{\dagger}L,\tilde{\rho}_{j}^{M,M^{\prime}}\right\}
+L\rho_{j}^{M,M^{\prime}}L^{\dagger}\nonumber\\
&+& [\tilde{\rho}_{j}^{M-1,M^{\prime}},L^{\dagger}]\xi_{j}
+[L, \tilde{\rho}_{j}^{M,M^{\prime}-1}]\xi^{\ast}_{j}\Big) \nonumber\\
&+& \bigg\{\frac{1}{k_{j}}\left(
L\tilde{\rho}_{j}^{M,M^{\prime}}L^{\dagger}+L\tilde{\rho}_{j}^{M,M^{\prime}-1}\xi^{\ast}_{j}
+\tilde{\rho}_{j}^{M-1,M^{\prime}}L^{\dagger}\xi_{j}\right.\nonumber\\
&+&\left.\tilde{\rho}_{j}^{M-1,M^{\prime}-1}|\xi_{j}|^2\right)
-\tilde{\rho}_{j}^{M,M^{\prime}}\bigg\}(\Delta n_{j}-k_{j}\tau).
\end{eqnarray}
The initial conditions to them are given by (\ref{ini}). In Eqs. (\ref{filteri}) and (\ref{filterj}) all terms that do not give contribution to the continuous case when $\tau \to dt$ are omitted. From the fact that $\tilde{\rho}_{j}^{M^{\prime},M}=\left(\tilde{\rho}_{j}^{M,M^{\prime}}\right)^{\dagger}$ it follows that there are at most $(N+1)(N+2)/2$ independent equations.

When the results of the measurement are not read out then the state of $\mathcal{S}$ at time $j\tau$ for $j>0$ is the {\it a priori} state
\begin{equation}
\sigma_{j}=\langle \tilde{\rho}_{j}\rangle_{st},
\end{equation}
being the mean value of $\tilde{\rho}_{j}$ taken with respect to the measure associated to (\ref{probtra}). Of course, initially $\sigma_{j=0}=|\psi\rangle\langle\psi|$. For the {\it a priori} state $\sigma_{j}$ we obtain from Eq. (\ref{filteri}) the difference equation
\begin{eqnarray}\label{master}
{\sigma}_{j+1}&=& {\sigma}_{j}+\tau\Big(-i[H_{\mathcal{S}},{\sigma}_{j}]-\frac{1}{2}\left\{L^{\dagger}L,{\sigma}_{j}\right\}
+L\sigma_{j}L^{\dagger}\nonumber\\
&+& [{\sigma}_{j}^{N-1,N},L^{\dagger}]\xi_{j}
+[L, {\sigma}_{j}^{N,N-1}]\xi^{\ast}_{j}\Big).
\end{eqnarray}
And for the operators
\begin{equation}
\sigma_{j}^{M,M^{\prime}}=\langle \tilde{\rho}_{j}^{M,M^{\prime}}\rangle_{st}
\end{equation}
we get the set of difference equations
\begin{eqnarray}
{\sigma}_{j+1}^{M,M^{\prime}}&=& {\sigma}_{j}^{M,M^{\prime}}+\tau\Big(-i[H_{\mathcal{S}},{\sigma}_{j}^{M,M^{\prime}}]-\frac{1}{2}\left\{L^{\dagger}L,{\sigma}_{j}^{M,M^{\prime}}\right\}
+L\sigma_{j}^{M,M^{\prime}}L^{\dagger}\nonumber\\
&+& [{\sigma}_{j}^{M-1,M^{\prime}},L^{\dagger}]\xi_{j}
+[L,{\sigma}_{j}^{M,M^{\prime}-1}]\xi^{\ast}_{j}\Big)
\end{eqnarray}
with the initial conditions  ${\sigma}_{j=0}^{M,M^{\prime}}=N!/\sqrt{M!M^{\prime}!}\delta_{MM^{\prime}}|\psi\rangle\langle\psi|$, {where} $0\leq M\leq N$ and $0\leq M^{\prime}\leq N$. Clearly, $\sigma^{N,N}_{j}=\sigma_{j}$.

\subsection{Differential stochastic master equations}

To obtain the continuous in time evolution of $\mathcal{S}$ we fix time $t=j\tau$, where
$j$ is the number of the bath elements which interacted with $\mathcal{S}$ up to $t$. Note that the time $t$ is fixed but arbitrary and when $j\to+\infty$ we have $\tau\to 0$. In {the continuous time limit} from (\ref{filterj}) we obtain the stochastic differential equations
\begin{eqnarray}\label{filterl}
d\tilde{\rho}_{t}^{M,M^{\prime}}&=& dt\Big(-i[H_{\mathcal{S}},\tilde{\rho}_{t}^{M,M^{\prime}}]-\frac{1}{2}\left\{L^{\dagger}L,\tilde{\rho}_{t}^{M,M^{\prime}}\right\}
+L\rho_{t}^{M,M^{\prime}}L^{\dagger}\nonumber\\
&+& [\tilde{\rho}_{t}^{M-1,M^{\prime}},L^{\dagger}]\xi_{t}
+[L, \tilde{\rho}_{t}^{M,M^{\prime}-1}]\xi^{\ast}_{t}\Big) \nonumber\\
&+& \bigg\{\frac{1}{k_{t}}\left(
L\tilde{\rho}_{t}^{M,M^{\prime}}L^{\dagger}+L\tilde{\rho}_{t}^{M,M^{\prime}-1}\xi^{\ast}_{t}
+\tilde{\rho}_{t}^{M-1,M^{\prime}}L^{\dagger}\xi_{t}\right.\nonumber\\
&+&\left.\tilde{\rho}_{t}^{M-1,M^{\prime}-1}|\xi_{t}|^2\right)
-\tilde{\rho}_{t}^{M,M^{\prime}}\bigg\}\left(d n_{t}-k_{t}dt\right),
\end{eqnarray}
where
\begin{equation}\label{intensity2}
k_{t} =\mathrm{Tr}_{\mathcal{S}}\left( L^{\dagger}L\tilde{\rho}_{t}
+ \xi_{t}^{\ast}L\tilde{\rho}_{t}^{N,N-1}+ \xi_{t}
\tilde{\rho}_{t}^{N-1,N}L^{\dagger}+ |\xi_{t}|^2\tilde{\rho}_{t}^{N-1,N-1} \right),
\end{equation}
$\xi_{t}\in \mathbb{C}$ is the continuous version of $\xi_{j}$ satisfying the normalization condition
\begin{equation}
\int_{0}^{\infty}|\xi_{t}|^2dt=1,
\end{equation}
and initially
\begin{equation}
\tilde{\rho}_{t=0}^{M,M^{\prime}}=\frac{N!}{\sqrt{M!M^{\prime}!}}\delta_{M,M^{\prime}}\,|\psi\rangle\langle\psi|.
\end{equation}
By $n_{t}$ we {denote} the counting process defined by $\lim_{j\to+\infty}n_{j}$. One can check that the increment $dn_{t}=n_{t+dt}-n_{t}$ satisfies the relation $(dn_{t})^2=dn_{t}$ and the conditional mean value
\begin{equation}
\mathbbm{E}[dn_{t}|\tilde{\rho}_{t}]=k_{t}dt.
\end{equation}
From Eq. (\ref{filteri}) we get for the {\it a posteriori} state $\tilde{\rho}_{t}$ of $\mathcal{S}$ the differential stochastic equation of the form
\begin{eqnarray}\label{filterk}
d\tilde{\rho}_{t}&=& dt\Big(-i[H_{\mathcal{S}},\tilde{\rho}_{t}]-\frac{1}{2}\left\{L^{\dagger}L,\tilde{\rho}_{t}\right\}
+L\rho_{t}L^{\dagger}\nonumber\\
&+& [\tilde{\rho}_{t}^{N-1,N},L^{\dagger}]\xi_{t}
+[L, \tilde{\rho}_{t}^{N,N-1}]\xi^{\ast}_{t}\Big) \nonumber\\
&+& \bigg\{\frac{1}{k_{t}}\left(
L\tilde{\rho}_{t}L^{\dagger}+L\tilde{\rho}_{t}^{N,N-1}\xi^{\ast}_{t}+\tilde{\rho}_{t}^{N-1,N}L^{\dagger}\xi_{t}\right.\nonumber\\
&+&\left.\tilde{\rho}_{t}^{N-1,N-1}|\xi_{t}|^2\right)
-\tilde{\rho}_{t}\bigg\}(d n_{t}-k_{t}dt)
\end{eqnarray}
with the initial condition $\tilde{\rho}_{t=0}=|\psi\rangle\langle\psi|$.
Of course, $\tilde{\rho}_{t}=\tilde{\rho}_{t}^{N,N}$, so the stochastic master equation (\ref{filterk}) is one of (\ref{filterl}). The set of equations (\ref{filterl}) are equivalent to the set of SMEs (48) determined in \cite{BC17}.

Clearly, for the non-selective measurement we get from Eqs. (\ref{filterl}) and (\ref{filterk}) respectively
\begin{eqnarray}\label{master2}
\frac{d{\sigma}_{t}^{M,M^{\prime}}}{dt}&=&-i[H_{\mathcal{S}},{\sigma}_{t}^{M,M^{\prime}}]-\frac{1}{2}\left\{L^{\dagger}L,{\sigma}_{t}^{M,M^{\prime}}\right\}
+L\sigma_{t}^{M,M^{\prime}}L^{\dagger}\nonumber\\
&+& [{\sigma}_{t}^{M-1,M^{\prime}},L^{\dagger}]\xi_{t}
+[L,{\sigma}_{t}^{M,M^{\prime}-1}]\xi^{\ast}_{t}
\end{eqnarray}
and
\begin{eqnarray}\label{master1}
\frac{d{\sigma}_{t}}{dt}&=& -i[H_{\mathcal{S}},{\sigma}_{t}]-\frac{1}{2}\left\{L^{\dagger}L,{\sigma}_{t}\right\}
+L\sigma_{t}L^{\dagger}\nonumber\\
&+& [{\sigma}_{t}^{N-1,N},L^{\dagger}]\xi_{t}
+[L, {\sigma}_{t}^{N,N-1}]\xi^{\ast}_{t},
\end{eqnarray}
where $\sigma_{t}=\langle \tilde{\rho}_{t}\rangle_{st}$, $\sigma_{t}^{M,M^{\prime}}=\langle \tilde{\rho}_{t}^{M,M^{\prime}}\rangle_{st}$ and initially ${\sigma}_{t=0}^{M,M^{\prime}}=N!/\sqrt{M!M^{\prime}!}\delta_{MM^{\prime}}|\psi\rangle\langle\psi|$, where $0\leq M\leq N$ and $0\leq M^{\prime}\leq N$.

\section{Solution to master equation and photon statistics}

In this Section we return to notation with explicitly written condition for the results of the past measurement.

\subsection{Discrete case}

The solutions to Eqs. (\ref{rec2a}) and  (\ref{rec2b}) respectively are
\begin{equation}
|\psi^{N}_{j|\pmb{\eta}_{j}}\rangle=V_{\eta_{j} 0} V_{\eta_{j-1} 0} \dots V_{\eta_1 0} |\psi\rangle
\end{equation}
and for all $j\geq N-M$
\begin{equation}
|\psi^M_{j|\pmb{\eta}_{j}}\rangle = \sqrt{\frac{N!}{M!}}\sum_{\pmb{ r} \in \mathbb{N}^j : \sum_{k} r_{k} = N-M} \prod_{k=0}^{\stackrel{\longleftarrow}{j-1}}\sqrt{\tau}^{r_{k}} \xi_{k}^{r_{k}} V_{\eta_{k+1} r_{k}} |{\psi}\rangle,
\end{equation}
where $0\leq M\leq N-1$, the vector $\pmb{r}$ consists of zeros and ones, and the arrow means that we use the time ordered products. One can check that $|\psi^M_{j|\pmb{\eta}_{j}}\rangle =0$ for all $j<N-M$. Note that instead of the notation with the full vector $\pmb{\eta}_j$ we may provide the location of `$1$' in the string $(\eta_j,\ldots,\eta_1)$, that is $(l_s,\ldots,l_1)$ means that we observed $s$ photons at times $\tau_i = \tau l_i$ $(i=1,\ldots,s)$ and no other photons in the period from $0$ to $j\tau$. Of course, any $l_{i}\geq 1$. Thus, for observing no counts from $0$ to $j\tau$ we get
\begin{equation}\label{sol1}
|\psi^{N}_{j|0}\rangle=V_{00}^{j}|\psi\rangle
\end{equation}
and for all $j\geq M$ where $1\leq M\leq N$ we have
\begin{equation}\label{sol2}
|\psi^{N-M}_{j|0}\rangle= \sqrt{\frac{N!}{(N-M)!}}V_{00}^{j} \sum_{k_{M}=M-1}^{j-1}\sum_{k_{M-1}=M-2}^{k_{M}-M+1}\ldots\sum_{k_{1}=0}^{k_{2}-1}
W_{k_{M}}W_{k_{M-1}}\ldots W_{k_{1}}|\psi\rangle,
\end{equation}
where
\begin{equation}
W_{k}=V_{00}^{-k-1}\sqrt{\tau}\xi_{k}V_{01}V_{00}^{k}.
\end{equation}
Of course, $|\psi^{N-M}_{j|0}\rangle=0$ for $j<M$.

The formula (\ref{sol2}) has a simple interpretation. The vector $|\psi^{N-M}_{j|0}\rangle$ is associated with the scenario that the future part of the environment stores $N-M$ photons. When no photons were observed up to $j\tau$ it means that $M$ photons were simply absorbed by $\mathcal{S}$.

For a count at $l_{1}\tau$ and not other counts from $0$ to $j\tau$ we have
\begin{equation}\label{sol3}
|\psi_{j|l_{1}}^{N}\rangle=
V_{00}^{j-l_{1}}V_{10}V_{00}^{l_{1}-1}|\psi\rangle,
\end{equation}
for $j\geq 1$
\begin{eqnarray}\label{sol4}
|\psi_{j|l_1}^{N-1}\rangle&=& \sqrt{\tau N}
\left[V_{00}^{j-l_{1}}\xi_{l_{1}-1}V_{11}V_{00}^{l_{1}-1} \right. \nonumber \\ &+&
V_{00}^{j-l_{1}}V_{10}V_{00}^{l_{1}-1}\sum_{k=0}^{l_{1}-2}V_{00}^{-k-1}
\xi_{k}V_{01}V_{00}^k \\
&+&\left. \sum_{k=l_{1}}^{j-1}V_{00}^{j-k-1}\xi_{k}V_{01}V_{00}^{k-l_{1}}V_{10}V_{00}^{l_{1}-1}
\right]|\psi\rangle, \nonumber
\end{eqnarray}
for $j\geq 2$
\begin{eqnarray}\label{sol5}
|\psi_{j|l_1}^{N-2}\rangle&=& \sqrt{N(N-1)}\tau\left[V_{00}^{j-l_{1}}\xi_{l_{1}-1}V_{11}\sum_{k=0}^{l_{1}-2}V_{00}^{l_{1}-k-2} \xi_{k}V_{01}V_{00}^{k}\right.\nonumber\\
&+&\sum_{k=l_{1}}^{j-1}V_{00}^{j-k-1}\xi_{k}V_{01}V_{00}^{k-l_{1}}
\xi_{l_{1}-1}V_{11}V_{00}^{l_{1}-1}\nonumber\\
&+&V_{00}^{j-l_{1}}V_{10}
\sum_{k_{2}=1}^{l_{1}-2}\sum_{k_{1}=0}^{k_{2}-1}V_{00}^{l_{1}-k_{2}-2}
\xi_{k_{2}}V_{01}V_{00}^{k_{2}-k_{1}-1}\xi_{k_{1}}V_{01}V_{00}^{k_{1}}\nonumber\\
&+&\sum_{k_{2}=l_{1}+1}^{j-1}\sum_{k_{1}=l_{1}}^{k_{2}-1}V_{00}^{j-k_{2}-1}
\xi_{k_{2}}V_{01}V_{00}^{k_{2}-k_{1}-1}\xi_{k_{1}}V_{01}V_{00}^{k_{1}-l_{1}}V_{10}V_{00}^{l_{1}-1}\nonumber\\
&+&\left.\sum_{k_{2}=l_{1}}^{j-1}V_{00}^{j-k_{2}-1}\xi_{k_{2}}V_{01}V_{00}^{k_{2}-l_{1}}V_{10}
\sum_{k_{1}=0}^{l_{1}-2}V_{00}^{l_{1}-k_{1}-2}\xi_{k_{1}}V_{01}V_{00}^{k_{1}}
\right]|\psi\rangle,
\end{eqnarray}
and so on.

The formula (\ref{sol3}) shows that if the environment stores $N$ photons in the future, then the detected photon came from $\mathcal{S}$. The successive formulae become more and more involved, but their structures and interpretations are rather straightforward. One can recognize there two sources of the detected photons: the external field and the system $\mathcal{S}$. In the formula (\ref{sol4}) the successive terms are associated with the following scenarios: i) the detected photon came directly from the field, ii) first $\mathcal{S}$ absorbed a photon from the field and then $\mathcal{S}$ emitted a photon, iii) first $\mathcal{S}$ emitted a photon and then $\mathcal{S}$ absorbed some photon from the field.  Of course, for the particular system $\mathcal{S}$ and initial conditions
not all of these possibilities give non-zero contribution to $|\psi_{j|l_1}^{N-1}\rangle$.

Note that armed with the explicit forms for vectors $|\psi_{j|l_{s},\ldots,l_{1}}^{M}\rangle$ and $|\psi_{j|0}^{M}\rangle$ (where zero means that we do not observed any photons from $0$ to $j\tau$) for $0\leq M\leq N$ we can find the solution to the difference master equation (\ref{master}) and the whole statistics of the output photons. Namely, the solution to (\ref{master}) we can write in the form
\begin{equation}\label{mastersol}
\sigma_{j}=\rho_{j|0}+\sum_{s=1}^{j}\sum_{l_{s}=s}^{j}\sum_{l_{s-1}=s-1}^{j-1}\ldots\sum_{l_{1}=1}^{l_{2}-1}\rho_{j|l_{s},\ldots,l_{2},l_{1}},
\end{equation}
where
\begin{equation}
\rho_{j|0}=\sum_{M=0}^{N} p_j^M\, |\psi^{M}_{j|0}\rangle\langle\psi_{j|0}^{M}|  
\end{equation}
and
\begin{equation}
\rho_{j|l_{s},\ldots,l_{2},l_{1}} = \sum_{M=0}^{N} p_j^M\, |\psi^{M}_{j|l_{s},\ldots,l_{2},l_{1}}\rangle\langle\psi_{j|l_{s},\ldots,l_{2},l_{1}}^{M}| 
\end{equation}
Clearly, the quantities $\rho_{j|cond}$ are the unnormalized conditional density operators (\ref{cond4}). We sum in (\ref{mastersol}) over all possible pathways of detection from $s=0$ to $s=j$ photons in the interval from $0$ to $j\tau$.
The probability of registering no photons in the interval from $0$ to $j\tau$ is given by
\begin{equation}\label{prob2}
P_{0}^{j}(0)=\mathrm{Tr}_{\mathcal{S}}\rho_{j|0},
\end{equation}
where
\begin{equation}\label{prob3}
\mathrm{Tr}_{\mathcal{S}}\rho_{j|0}=\sum_{M=0}^{N} p_j^M \, \langle\psi^{M}_{j|0}|\psi^{M}_{j|0}\rangle . 
\end{equation}
The probability of registering $s$ photons from $0$ to $j\tau$ is
\begin{equation}
P_{0}^{j}(s)=\sum_{l_{s}=s}^{j}\sum_{l_{s-1}=s-1}^{j-1}\ldots\sum_{l_{1}=1}^{l_{2}-1}\mathrm{Tr}_{\mathcal{S}}\rho_{j|l_{s},\ldots,l_{2},l_{1}}
\end{equation}
where
\begin{equation}
\mathrm{Tr}_{\mathcal{S}}\rho_{j|l_{s},\ldots,l_{2},l_{1}} = \sum_{M=0}^{N} p_j^M\, \langle\psi^{M}_{j|l_{s},\ldots,l_{2},l_{1}}|\psi_{j|l_{s},\ldots,l_{2},l_{1}}^{M}\rangle 
\end{equation}

\subsection{Continuous case}

All realization of the counting process $n_{t}$ defined by Sec. 4.2 may be divided into disjoint sectors: $\mathcal{C}_s$ contains realizations with exactly $s$ counts in the interval from {time} $0$ to {time} $t$, one in each of the nonoverlapping intervals $[t_{1},t_{1}+dt_{1})$, $[t_{2},t_{2}+dt_{2})$, $\ldots$, $[t_{s},t_{s}+dt_{s})$, where $t_1 < t_{2}<\ldots < t_s$. Our goal is to find the formula for $|\psi^{N-M}_{t|t_s\ldots t_1}\rangle $. Since the general formula is rather involved we propose a simple diagrammatic representation. Let us introduce the following elementary processes and the corresponding diagrammatic representations:

\begin{enumerate}
	\item  free propagation up to time $t$
	
	\begin{equation}\label{}
	\mathbf{T}_t = e^{-iGt} \ \ \longleftrightarrow \ \ {----},
	\end{equation}
	where $G=H_{\mathcal{S}}-\frac{i}{2}L^{\dagger}L ,$ is a non-Hermitian Hamiltonian (like in the Wigner-Weisskopf theory),

	\item absorption of a photon by the system $\mathcal{S}$ from the environment at time $t$
	
	\begin{equation}
	W_{t}=-\mathbf{T}_{-t}\xi_{t}L^{\dagger}\mathbf{T}_{t} \ \ \longleftrightarrow \ \  --{\bullet}--,
	\end{equation}
	
	\item emission of a photon by the systems $\mathcal{S}$ to the detector at time $t$
	
	\begin{equation}\label{}
	\widetilde{L}_t = \mathbf{T}_{-t}L\mathbf{T}_{t} \ \ \longleftrightarrow \ \ --{\circ}--,
	\end{equation}
	
	\item absorption of a photon by a detector from the environment  at time $t$
	
	\begin{equation}\label{}
	\xi_t \ \ \longleftrightarrow \ \ \ast.
	\end{equation}
	
\end{enumerate}

Now, in the representation corresponding to $|\psi^{N-M}_{t|t_s\ldots t_1}\rangle $ one has the following ``Feynman rules'':

$$   {\rm Number\ of}\ (\ast + \circ) = s, $$

$$   {\rm Number\  of}\ (\ast + \bullet) = M .$$

Let us illustrate this technique by a few examples:

\begin{enumerate}

	\item for zero counts from {time} $0$ to {time} $t$ we obtain in the continuous limit from (\ref{sol1}) the formula
	\begin{equation}\label{zeroa}
	|\psi_{t|0}^{N}\rangle = \mathbf{T}_t |\psi\rangle
	\end{equation}
	and the corresponding trivial diagram: $\    {-----}\ $,
	
	\item  from (\ref{sol2}) one can find
	\begin{equation}
	|\psi^{N-M}_{t|0}\rangle=  
	\sqrt{ \frac{N\, !}{(N-M)\,!}}\,	\mathbf{T}_{t} \, \int_{0}^{t}dt_{M} \ldots\int_{0}^{t_{3}}dt_{2}\int_{0}^{t_{2}}
	dt_{1} W_{t_{M}}\ldots W_{t_{2}}W_{t_{1}}|\psi\rangle,
	\end{equation}
	with the corresponding diagram: $\   --{\bullet}--{\bullet}-- \ldots --{\bullet}--\ $,
	with exactly $M$ elementary processes $--{\bullet}--$,
	
	\item for a count at the time $t^{\prime}$ and no other counts in the interval $(0,t]$ we have the conditional vectors:
	\begin{enumerate}
		\item from (\ref{sol3})
		
		\begin{equation}
		|\psi_{t|t^{\prime}}^{N}\rangle = \sqrt{dt^{\prime}}\mathbf{T}_{t-t^{\prime}} L \mathbf{T}_{t^{\prime}} |\psi\rangle = \sqrt{dt^{\prime}}\mathbf{T}_{t} \,  \widetilde{L}_{t'} |\psi\rangle ,
		\end{equation}
		and the diagram: $\ --{\circ}-- \ $,
		
		\item from (\ref{sol4})
		\begin{eqnarray}
		|\psi_{t|t^{\prime}}^{N-1}\rangle &=&  \sqrt{Ndt^{\prime}}\Big[  \xi_{t^{\prime}} - \mathbf{T}_{t-t'} L\int_{0}^{t^{\prime}}ds \, \mathbf{T}_{t'-s} \xi_{s} L^{\dagger} \mathbf{T}_s  \nonumber \\
		&-& \int_{t^{\prime}}^{t} ds \mathbf{T}_{t-s} \xi_{s} L^{\dagger} \mathbf{T}_{s-t^{\prime}} L \mathbf{T}_{t^{\prime}}
		\Big]|\psi\rangle \\
		&=&   \sqrt{Ndt^{\prime}}\,  \mathbf{T}_t \Big[ \xi_{t^{\prime}} +  \widetilde{L}_{t'} \int_0^{t'} ds W_s + \int_{t'}^{t} ds W_s \widetilde{L}_{t'} \Big]|\psi\rangle \nonumber
		\end{eqnarray}
		and the diagram: $\   -\ast-\ +\ -\circ-\bullet-\ +\ -\bullet-\circ-\ $,

		\item from (\ref{sol5})
		\begin{eqnarray}\label{veccon2}
		|\psi_{t|t^{\prime}}^{N-2}\rangle &=&  \sqrt{N(N-1)dt^{\prime}}\Big[  \mathbf{T}_{t-t^{\prime}}L \int_{0}^{t^{\prime}}dt_{2}\int_{0}^{t_{2}}dt_{1}\mathbf{T}_{t^{\prime}-t_{2}}\xi_{t_{2}}L^{\dagger}\mathbf{T}_{t_{2}-t_{1}}
		\xi_{t_{1}}L^{\dagger}\mathbf{T}_{t_{1}}\nonumber\\
		&+&\int_{t^{\prime}}^{t}dt_{2}\int_{t^{\prime}}^{t_{2}}dt_{1}\mathbf{T}_{t-t_{2}}\xi_{t_{2}}L^{\dagger}\mathbf{T}_{t_{2}-t_{1}}\xi_{t_{1}}
		L^{\dagger}\mathbf{T}_{t_{1}-t^{\prime}}L\mathbf{T}_{t^{\prime}}
		\nonumber\\
		&+& \int_{t^{\prime}}^{t}dt_{2}\mathbf{T}_{t-t_{2}}\xi_{t_{2}}L^{\dagger}\mathbf{T}_{t_{2}-t^{\prime}}L\int_{0}^{t^{\prime}}dt_{1}\mathbf{T}_{t^{\prime}-
			t_{1}}L^{\dagger}\mathbf{T}_{t_{1}}\nonumber\\
		&-&\mathbf{T}_{t-t^{\prime}}\xi_{t^{\prime}}\int_{0}^{t^{\prime}}dt_{1}\mathbf{T}_{t^{\prime}
			-t_{1}}\xi_{t_{1}}L^{\dagger}\mathbf{T}_{t_{1}}-
		\int_{t^{\prime}}^{t}dt_{1}\mathbf{T}_{t-t_{1}}\xi_{t_{1}}L^{\dagger}\mathbf{T}_{t_{1} -t^{\prime}}\xi_{t^{\prime}}\mathbf{T}_{t^{\prime}}\Big]|\psi\rangle.  \\
		&=& \sqrt{N(N-1)dt^{\prime}}\, \mathbf{T}_t \, \Big[ \widetilde{L}_{t'} \int_{0}^{t^{\prime}}dt_{2}\int_{0}^{t_{2}}dt_{1} W_{t_2} W_{t_1} +
		\int_{t^{\prime}}^{t}dt_{2}\int_{t^{\prime}}^{t_{2}}dt_{1} W_{t_2} W_{t_1}\,  \widetilde{L}_{t'} \nonumber \\
		&+& \int_{t^{\prime}}^{t}dt_{2} W_{t_2} \, \widetilde{L}_{t'} \, \int_{0}^{t^{\prime}}dt_{1} W_{t_1} + \xi_{t'} \int_{0}^{t^{\prime}}dt_{1} W_{t_1} +  \int_{t'}^{t}dt_{1} W_{t_1}\, \xi_{t'} \Big]|\psi\rangle
		\end{eqnarray}
		
		and the diagram:
		
		$$   -\circ-\bullet-\bullet- \ +  \ -\bullet-\bullet-\circ-\ + \ -\bullet-\circ-\bullet- \ + \ -\ast-\bullet-\ + \ -\bullet-\ast-\,. $$

	\end{enumerate}

\end{enumerate}

The solution to Eq. (\ref{master1}) can written in the form
\begin{equation}
\sigma_{t}=\rho_{t|0}+\sum_{s=1}^{+\infty}\int_{0}^{t}dt_{s}\int_{0}^{t_{s}}dt_{s-1}\ldots
\int_{0}^{t_{2}}dt_{1}\rho_{t|t_{s},t_{s-1},\ldots,t_{2},t_{1}}
\end{equation}
where
\begin{equation}
\rho_{t|0}=\sum_{M=0}^{N} p_t^M\, |\psi^{M}_{t|0}\rangle\langle\psi^{M}_{t|0}| , 
\end{equation}
with
\begin{equation}\label{}
p_t := \int_{t}^{+\infty}dt^{\prime}|\xi_{t^{\prime}}|^2 ,
\end{equation} 
and
\begin{eqnarray}
\lefteqn{dt_{s}dt_{s-1}\ldots dt_{1}\rho_{t|t_{s},t_{s-1},\ldots,t_{2},t_{1}}=}
\\&&\sum_{M=0}^{N} p_t^M\, |\psi^{M}_{t|t_{s},t_{s-1},\ldots,t_{2},t_{1}}\rangle\langle\psi^{M}_{t|t_{s},t_{s-1},\ldots,t_{2},t_{1}}|
\end{eqnarray}
with the initial condition $|\psi^{M}_{j=0}\rangle=\delta_{NM}|\psi\rangle$ for $0\leq M\leq N$.
The expression under the integrals can be interpreted as the unnormalized conditioned density operator of $\mathcal{S}$. We take the integrals over all possible realization of the stochastic process $n_{t}$.

{We can easily write down the formulae for the {\it a priori} (unconditional) probabilities characterizing the counting process $n_{t}$.} Namely, the probability of no counts up to time $t$ is given as
\begin{equation}
P_{0}^{t}(0)=\sum_{M=0}^{N} p_t^M \, \langle\psi^{M}_{t|0}|\psi^{M}_{t|0}\rangle .
\end{equation}
The probability density $ p_{0}^{t}(t_{s}, t_{s-1}, \ldots, t_{2}, t_{1})$ of observing a particular trajectory corresponding to $s$ counts in the interval from $0$ to $t$, one in each of the nonoverlapping intervals $[t_{1},t_{1}+dt_{1})$, $[t_{2},t_{2}+dt_{2})$, $\ldots$, $[t_{s},t_{s}+dt_{s})$, where $t_1 < t_{2}<\ldots < t_s$ is defined by
\begin{eqnarray}
\lefteqn{ p_{0}^{t}(t_{s}, t_{s-1}, \ldots, t_{2}, t_{1}){dt_{s}dt_{s-1}\ldots dt_{1}} =}
\\&& \sum_{M=0}^{N} p_t^M\, \langle\psi_{t|t_{s}, t_{s-1}, \ldots, t_{2}, t_{1}}^{M}|\psi_{t|t_{s}, t_{s-1}, \ldots, t_{2}, t_{1}}^{M}\rangle \nonumber 
\end{eqnarray}
Of course, the probability of having exactly $s$ counts up to time $t$ reads
\begin{equation}
P_{0}^{t}(s)\!=\!\int_{0}^{t}\!dt_{s}\!\int_{0}^{t_{s}}\!dt_{s-1}\!\ldots\!\int_{0}^{t_{2}}\!dt_{1}
p_{0}^{t}(t_{s}, \!t_{s-1}, \!\ldots \!, \!t_{2}, t_{1}) .
\end{equation}
Moreover, the mean time of $s$-th count is defined by
\begin{equation}
\overline{t}_{s}=\int_{0}^{+\infty}dt_{s}t_{s}\int_{0}^{t_{s}}dt_{s-1}\ldots \int_{0}^{t_{2}}dt_{1}p(t_{s},t_{s-1},\ldots,t_{2},t_{1}),
\end{equation}
Clearly, the mean time of the entanglement between the system $\mathcal{S}$ and the environment is $\mathrm{max}_{s\in \mathbbm{N}}$$\overline{t}_{s}$.

\section{Conclusions}

{In this paper we have derived the set of SMEs for a quantum system interacting with the bosonic field prepared in a continuous-mode $N$-photon state. The filtering equations, describing the continuous in time conditional evolution of an open quantum system, have been determined for the photon counting detection of the output field. These equations have been obtained as the limit of difference equations determined with making use of a collision model \cite{C17}. The temporal correlations present in the input field imply that the system becomes entangled with the environment and that the evolution of the open system is non-Markovian. We have shown that the quantum system interacting with the field in $N$-photon state can emit or absorb at most one photon at a given moment, which is an important feature of the problem, discussed also, for instance, in \cite{RS16}. Our results are in a perfect agreement with the earlier studies performed with making use of QSC \cite{BCBC12,SZX16,BC17}. We would like to emphasize that our approach allows not only to determine the differential equations for the conditional as well as unconditional state of the system but it also enables one to find the general structure of quantum trajectories together with a clear physical interpretation. In the continuous time limit the solutions to the SME are represented by a simple diagrammatic technique with very transparent ``Feynman rules''. This diagrammatic technique considerably simplifies the structure of the solutions and enables one to find physical interpretation to them in terms of a few elementary processes. We have derived the probability density for the observing counting process and we have given an explicit formula for the {\it a priori} state with making use of the unnormalized conditional density operators.}

In a forthcoming paper we plan to provide the detailed analysis of various non-Markovian effects and compare our framework with well established approaches like divisibility of the evolution or memory kernel approach.

\appendix

\section{Temporal decomposition of $n$-photon state}\label{A}

To find a division of $|N_{\xi}\rangle$ into the vectors of the Hilbert spaces $\mathcal{H}_{\mathcal{E}}^{l]}$ and
$\mathcal{H}_{\mathcal{E}}^{[l+1}$, we use the following property
\begin{equation}\label{div}
\left(b_{\xi}^{\dagger}\right)^{N}=\sum_{M=0}^{N} {N \choose M}\left(A^{l]}\right)^{M} \left(B^{[l+1}\right)^{N-M},
\end{equation}
where
\begin{equation}
A^{l]}=\sum_{k=0}^{l} \sqrt{\tau}\xi_{k}\mathbf{b}_{k}^{\dagger},
\end{equation}
\begin{equation}
B^{[l+1}=\sum_{k=l+1}^{+\infty}\sqrt{\tau}\xi_{k}\mathbf{b}_{k}^{\dagger},
\end{equation}
and
\begin{equation}
{N \choose M}=\frac{N!}{M!\left(N-M\right)!}.
\end{equation}
Note that the operator $A^{l]}$ acts non-trivially only in the Hilbert space $\mathcal{H}_{\mathcal{E}}^{l]}$ and $B^{[l+1}$ acts non-trivially only in $\mathcal{H}_{\mathcal{E}}^{[l+1}$, so they both commute.
Hence, we get
\begin{equation}
|N_{\xi}\rangle =\sum_{M=0}^{N} \sqrt{N \choose M}|M_{\xi}\rangle_{[0,l]}\otimes|\left(N-M\right)_{\xi}\rangle_{[l+1,+\infty)},
\end{equation}
where
\begin{equation}
|M_{\xi}\rangle_{[r,l]} = \frac{1}{\sqrt{M!}}\left(\sqrt{\tau}\xi_{r} b_{r}^{\dagger}\otimes\mathbbm{1}_\mathcal{E}^{[r+1,l]}+\sum_{k=r+1}^{l}\mathbbm{1}_\mathcal{E}^{[r,k-1]} \otimes \sqrt{\tau}\xi_{k} b_{k}^{\dagger}\otimes\mathbbm{1}_\mathcal{E}^{[k+1,l]}\right)^{M}|vac\rangle_{[r,l]}
\end{equation}
and $|vac\rangle_{[r,l]} =|0\rangle_{r}\otimes|0\rangle_{r+1}\otimes\ldots\otimes|0\rangle_{l}$. Of course, $|0_{\xi}\rangle_{[r,l]} \equiv|vac\rangle_{[r,l]} $. It is clear that $|M_{\xi}\rangle_{[r,l]}$ is the vector from the Hilbert space $\mathcal{H}_{\mathcal{E}}^{[r,l]}$ and one can check that
\begin{equation}\label{norm}
_{[r,l]}\langle M^{\prime}_{\xi}|M_{\xi}\rangle_{[r,l]}=\delta_{M^{\prime}M}\left(\sum_{k=r}^{l}\tau|\xi_{k}|^2\right)^M.
\end{equation}
Thus for any $l\geq 1$ we have
\begin{equation}\label{pcf}
|N_{\xi}\rangle =\sum_{M=0}^{N}\sum_{M^{\prime}=0}^{M} \sqrt{{N \choose M}{M \choose M^{\prime}}}\left(\sqrt{\tau}\xi_{l}\right)^{M-M^{\prime}}|M^{\prime}_{\xi}\rangle_{[0,l-1]}\otimes|M-M^{\prime}\rangle_{l} \otimes|(N-M)_{\xi}\rangle_{[l+1,+\infty)},
\end{equation}
which can be interpreted as a division of the vector $|N_{\xi}\rangle$ into the past, current, and future vectors of $\mathcal{E}$. Moreover, by Eqs. (\ref{norm}) and (\ref{pcf}) one can check that for $|N_{\xi}\rangle$ the probability of detecting zero photons in the time interval $[l\tau,(l+1)\tau)$
\begin{equation}
P_{l}(0)=\langle N_{\xi}|\left(\mathbbm{1}_{\mathcal{E}}^{l-1]}\otimes|0\rangle_{l}\langle 0|_{l}\otimes\mathbbm{1}_{\mathcal{E}}^{[l+1}\right)|N_\xi\rangle=1+O(\tau),
\end{equation}
the probability of detecting one photon
\begin{equation}
P_{l}(1)=\langle N_{\xi}|\left(\mathbbm{1}_{\mathcal{E}}^{l-1]}\otimes|1\rangle_{l}\langle 1|_{l}\otimes\mathbbm{1}_{\mathcal{E}}^{[l+1}\right)|N_\xi\rangle=N\tau|\xi_{l}|^2+O(\tau^2),
\end{equation}
and the probability of detecting two photons
\begin{equation}
P_{l}(2)=\langle N_{\xi}|\left(\mathbbm{1}_{\mathcal{E}}^{l-1]}\otimes|2\rangle_{l}\langle 2|_{l}\otimes\mathbbm{1}_{\mathcal{E}}^{[l+1}\right)|N_\xi\rangle=O(\tau^2).
\end{equation}
We thus come to conclusion that if the environment is prepared in the $N$-photon state $|N_{\xi}\rangle$ and $\tau$ is treated as a small time step we can detect at most one photon at a given moment. The probability of detecting more that one photon is in this case $O(\tau^2)$ and it is negligible (see \cite{RB17}).

\section{Proof of Theorem}   \label{C}

We prove Theorem by induction. Let us assume that observe that (\ref{cond5}) is true. Observe that
\begin{equation}\label{}
|M_{\xi}\rangle_{[j+1,+\infty)} = \sum_{M^{\prime}=0}^{M}\sqrt{M \choose M^{\prime}} \left(\sqrt{\tau} \xi_{j}\right)^{M^{\prime}} |M^{\prime}\rangle_{j}\otimes |\left(M-M^{\prime}\right)_{\xi}\rangle_{[j+1,+\infty)}.
\end{equation}
Hence, we may rewrite (\ref{cond5}) as follows
\begin{equation}
|\Psi_{j|\pmb{\eta}_j}\rangle=
\sum_{M=0}^{N}\sum_{M^{\prime}=0}^{M}\sqrt{M \choose M^{\prime}} \left(\sqrt{\tau} \xi_{j}\right)^{M^{\prime}} |M^{\prime}\rangle_{j}\otimes|\left(M-M^{\prime}\right)_{\xi}\rangle_{[j+1,+\infty)}\otimes|\psi_{j|\pmb{\eta}_j}^{M}\rangle .
\end{equation}
Now, acting  by the unitary operator $\mathbbm{V}_{[j}$  on the vector $|\Psi_{j|\pmb{\eta}_j}\rangle$ one finds
\begin{equation}
\mathbbm{V}_{[j} \,|\Psi_{j|\pmb{\eta}_j}\rangle = \sum_{M=0}^N \sum_{M^{\prime}=0}^{M} \sqrt{M \choose M^{\prime}} \left(\sqrt{\tau} \xi_j\right)^{M^{\prime}} \sum_{R=0}^{+\infty} |R\rangle_{j} \otimes |(M-M^{\prime})_\xi\rangle_{[j+1,\infty)} \otimes V_{RM^{\prime}} |\psi_{j|\pmb{\eta}_j}^M\rangle.
\end{equation}
Taking into account that the conditional vector $|\Psi_{j+1|  \pmb{\eta}_{j+1}} \rangle$ from    $\mathcal{H}_{\mathcal{E}}^{[j+1}\otimes \mathcal{H}_{S}$ is defined by

\begin{equation}\label{measurement}
\left(\Pi_{\eta_{j+1}}\otimes\mathbbm{1}_{\mathcal{E}}^{[j+1}\otimes\mathbbm{1}_{S}\right) \,\mathbbm{V}_{[j}\,|\Psi_{j|  \pmb{\eta}_j} \rangle=|\eta_{j+1}\rangle_{j}\otimes|\Psi_{j+1|  \pmb{\eta}_{j+1}} \rangle,
\end{equation}
where $\Pi_{\eta_{j+1}}=|\eta_{j+1}\rangle_{j}\langle \eta_{j+1}|$, and $\eta_{j+1}\in \mathbb{N}$ is the random result of the measurement of (\ref{obs}) at the time $\tau(j+1)$, one gets

\begin{equation}
|\Psi_{j+1|\pmb{\eta}_{j+1}}\rangle = \sum_{M=0}^N \sum_{M^{\prime}=0}^{M} \sqrt{M \choose M^{\prime}} \left(\sqrt{\tau} \xi_j\right)^{M^{\prime}} |\left(M- M^{\prime}\right)_\xi\rangle_{[j+1,+\infty)} \otimes V_{\eta_{j+1} M^{\prime}} |\psi_{j|\pmb{\eta}_j}^M\rangle.
\end{equation}
In order to obtain the recurrence formulae (\ref{rec}) we need to change the index of summation introducing $R=M-M^{\prime}$ such that
\begin{equation}\label{cond2}
|\Psi_{j+1|\pmb{\eta}_{j+1}}\rangle = \sum_{R=0}^N \sum_{M^{\prime}=0}^{N-R} \sqrt{R+ M^{\prime} \choose  M^{\prime}} (\sqrt{\tau} \xi_j)^{ M^{\prime}}  |R_\xi\rangle_{[j+1,+\infty)} \otimes V_{\eta_{j+1} M^{\prime}} |\psi_{j|\pmb{\eta}_j}^{R+ M^{\prime}}\rangle,
\end{equation}
which end up the proof.

\section*{Acknowledgements}

This paper was partially supported by the National Science Center project 2015/17/B/ST2/02026.

\end{document}